\documentclass[aps,prd,draft,amsmath,amssymb,superscriptaddress,preprintnumbers,nofootinbib,a4paper,11pt]{revtex4}
\pdfoutput=1
\usepackage{amsfonts,amsmath,amssymb,mathrsfs}
\usepackage{hyperref}
\usepackage{subfigure}
\usepackage{color}
\usepackage{graphicx}  
\usepackage{dcolumn}   
\usepackage{bm}        
\usepackage[english]{babel}
\definecolor{cp3}{cmyk}{0,.88,.77,.40}
\usepackage[usenames,dvipsnames]{xcolor}
\def\bea{\begin{eqnarray}}
\def\eea{\end{eqnarray}}

\begin{document}

\title{\Large{\color{cp3}{Composite Inflation Setup and Glueball Inflation}}}

\author{\color{RedViolet}{Fedor Bezrukov}}\email{Fedor.Bezrukov@physik.uni-muenchen.de}
\affiliation{\em Arnold Sommerfeld Center for Theoretical Physics, Department f\"ur Physik,\\Ludwig-Maximilians-Universit\"at M\"unchen,\\Institute for Nuclear Research of the Russian Academy of Sciences, \\60th October Anniversary, prospect 7a \\Moscow 117312, Russia}
\author{\color{RedViolet}{Phongpichit Channuie}}\email{channuie@cp3.dias.sdu.dk}
\author{\color{RedViolet}{Jakob Jark J\o rgensen}}\email{joergensen@cp3.dias.sdu.dk}
 \author{\color{RedViolet}{and Francesco Sannino}}\email{sannino@cp3.dias.sdu.dk}
 \affiliation{{\color{cp3}{ \rm CP}$^{\bf 3}${\rm-Origins}} \& the Danish Institute for Advanced Study {\color{cp3}{{\rm DIAS}}},\\ University of Southern Denmark, Campusvej 55, \\DK-5230 Odense M, Denmark.}

\begin{abstract}
We explore the paradigm in which inflation is driven by a four-dimensional strongly coupled dynamics with a non-minimal coupling to gravity. We introduce a model where the inflaton is identified with the glueball field of a pure Yang-Mills theory.  We introduce the dilatonic-like glueball action, which is obtained by requiring saturation of the underlying Yang-Mills trace anomaly at the effective action level. We couple the resulting action non-minimally to gravity. We demonstrate that it is possible to achieve successful inflation with the confining scale of the underlying Yang-Mills theory naturally of the order of the grand unified energy scale. We also argue that the metric formulation gives a more consistent picture for models of composite inflation than the Palatini one.  Finally we  show that, within the metric formulation, the model nicely respects tree-level unitarity for the scattering of the inflaton field all the way to the Planck scale.\\ 
[1cm]

\end{abstract}

\maketitle

\section{Introduction} 
 
Two prominent physics problems, i.e. the origin of mass of all the standard model particles and inflation \cite{Starobinsky:1979ty,Starobinsky:1980te,Mukhanov:1981xt,Guth:1980zm,Linde:1981mu,Albrecht:1982wi}, the mechanism responsible for an early rapid expansion of our Universe, are both modeled traditionally via the introduction of new scalar fields. 
 
However, field theories featuring fundamental scalars are unnatural. The reason being that typically these theories lead to the introduction of symmetry-unprotected super-renormalizable operators, such as the scalar quadratic mass operator. Quantum corrections therefore introduce untamed divergencies which have to be fine-tuned away. Furthermore the basic description of space-time is via spinors and fundamental scalars have not yet been observed in nature. 

It is well known that new strong dynamics can replace the Higgs mechanism \cite{arXiv:0911.0931,arXiv:0804.0182}.  Another logical possibility is that theories with scalars are gauge-dual to theories featuring only fermionic degrees of freedom \cite{Sannino:2009qc,Sannino:2010fh,Sannino:2011mr,Mojaza:2011rw}. Recently we have also shown that it is possible to construct models in which the inflaton emerges as a composite state of a four-dimensional strongly coupled theory \cite{Channuie:2011rq}. In section \ref{Glueball}  of this work we consider a model where the inflaton emerges as the lightest glueball field associated to, in absence of gravity, a pure Yang-Mills theory. This theory constitutes the archetype of any composite model in flat space and consequently of models of composite inflation. We show that it is possible to achieve successful glueball inflation. Furthermore the natural scale of compositeness associated to the underlying Yang-Mills gauge theory, for the consistence of the model, turns to be of the order of the grand unified scale. This result is in agreement with the scale of compositeness scale determined in \cite{Channuie:2011rq} for a very different underlying model of composite inflation. We also argue that within the metric formulation models of composite inflation behave better than within the Palatini one.  In section \ref{unitarity} we investigate the tree-level unitarity constraints, for inflaton scattering, at the effective action level in the Einstein frame and for both the Palatini and metric formulation. We discover that the unitarity cutoff, i.e. the scale above which the model ceases to be valid and gravitational corrections must be taken into account, is nicely the Planck scale for the metric formulation while it is the strongly coupled Yang-Mills scale for the Palatini one. The metric formulation provides therefore a consistent picture for a successful glueball inflation model. We provide an extensive discussion of the effects of graviton scattering in section \ref{graviton-scattering}. We summarize the relevant energy scales of the problem in section \ref{scales}. We finally conclude in section \ref{conclusions}. 

In the appendix we generalize the paradigm introduced in \cite{Channuie:2011rq} by first spelling out the setup for generic models of composite inflation. Within this framework we determine useful expressions for the slow-roll parameters for composite inflation. 

\section{Glueball/Dilaton Inflation} 
\label{Glueball}

Pure Yang-Mills theories featuring only gluonic-type fields are the simplest examples of strongly coupled theories. It is therefore natural to investigate composite inflation using these theories. Who is then the inflaton? The candidate is the interpolating field describing the lightest glueball. 
\begin{equation}
\Phi(x) = \frac{\beta}{g} {\rm Tr} \left[ \mathcal{G}^{\mu \nu} \mathcal{G}_{\mu \nu}\right] \ ,
\end{equation}
where $\mathcal{G}^{\mu \nu}$ is the standard non-abelian field strength and $\beta$ is the full beta function of the theory in any renormalization scheme. $\Phi$ is written in a renormalization scheme-free way and therefore is associated to a physical quantity. 
The Yang-Mills trace anomaly constrains the low energy effective Lagrangian for the lightest gluebll state \cite{Schechter:1980ak,Migdal:1982jp,Cornwall:1983zb}to be: 
\begin{equation}
\mathcal{L}_{\rm GI} = \Phi^{-\frac{3}{2}}\partial_{\mu}\Phi \partial^{\mu} \Phi  - V_{ \rm GI}, \quad V_{\rm GI} =  \frac{\Phi}{2} \ln \left( \frac{\Phi}{\Lambda}\right)  \ . \label{effective}
\end{equation}

The generalization of this action, at the effective Lagrangian level, allowing also for a description of the topological properties of the theory  can be found here \cite{Sannino:1999qe,Hsu:1998jd}.  This generalization, and associated operators, by construction cannot affect the potential above nor the following analysis involving gravity. The reason being that the resulting action {\it must saturate} the underlying trace-anomaly only via the effective potential above. We discuss, however, the naive effects of higher order operators on graviton-scattering in section (\ref{graviton-scattering}). This low energy effective Lagrangian, at times, is also known as the action for the {\it dilaton}. This is so since the composite scalar field $\Phi$ saturates the dilatonic current. Therefore we could as well have called the model we are about to introduce, non-minimal dilaton inflation. In the future we plan also to investigate perturbative dilatonic actions \cite{Antipin:2011aa}. A recent use for the action above for the electroweak physics and {\it cosmology} can be found here \cite{Dietrich:2005jn,Campbell:2011iw} 

We consider the following coupling of $\Phi$ to gravity in the Jordan frame: 
\begin{align}
\mathcal{S}_{\rm CI,J}=\int d^{4}x \sqrt{-g}\left[-\frac{\mathcal{M}^{2}+\xi\,{\Phi}^{\frac{1}{2}}}{2}g^{\mu\nu}R_{\mu\nu}+\mathcal{L}_{\rm GI}\right]\ .
\end{align}
 In this framework $\mathcal{M}$ is not automatically the Planck constant $M_{\rm Pl}$. The non-minimal coupling to gravity is controlled by the dimensionless coupling $\xi$. The non-analytic power of $\Phi$  emerges because we are requiring a dimensionless coupling with the Ricci scalar.  

It is convenient to introduce the field $\varphi$ possessing unity canonical dimension and related to $\Phi$ as follows: 
\begin{equation}
\Phi = \varphi^4 \ .
\end{equation}
The non-minimally coupled glueball effective action to gravity then reads: 
\begin{align}
\mathcal{S}_{\rm GI,J}=\int d^{4}x \sqrt{-g}\left[-\frac{\mathcal{M}^2 +  \xi  \varphi^2}{2}g^{\mu\nu}R_{\mu\nu}  + 16  \partial_{\mu} \varphi \partial^{\mu} \varphi - V_{ \rm GI} \right] , \quad V_{\rm GI} =   2 \varphi^4 \ln \left( \frac{\varphi}{\Lambda}\right).  \label{gLH}
\end{align}

Imposing the conformal transformation with 
\begin{equation}
\Omega^2 = \frac{\mathcal{M}^2 + \xi \varphi ^2}{M^{2}_{\rm P}}  \ ,
\end{equation}
 the action in the Einstein frame reads: 
\begin{equation}
\mathcal{S}_{\rm GI}=\int d^{4}x \sqrt{-{g}}\left[-\frac{M_{\rm P}^{2}}{2}{g}^{\mu\nu}R_{\mu\nu}+16 \Omega^{-2}\left(1+\frac{3f\Omega^{-2}\xi^2 \varphi^2}{16M^{2}_{\rm P}}\right)g^{\mu\nu}\,\, \partial_{\mu}\varphi \partial_{\nu}\varphi-\Omega^{-4}V_{\rm GI}\right]. \label{Gnonminimal}
\end{equation}
{ Where $f=0$ corresponds to the Palatini formulation and $f=1$ to the metric case. See also the Appendix for relevant references on the difference between the Palatini and metric approach.}
We have left the explicit dependence on $\varphi$ rather than using the canonically normalized new scalar field $\chi = \chi(\varphi)$ introduced in the appendix (\ref{CISET}). 
We are now able to determine the slow-roll parameters and constraints relevant for inflation. From (\ref{epsilon}) we obtain in the large field regime: 
 \begin{equation}
 \varphi^2 \gg \frac{\mathcal{M}^2}{\xi} \ . 
 \label{lfe}
 \end{equation}
we derive the following slow-roll parameter $\epsilon$: 
\begin{align}
\epsilon \simeq 
 \frac{1}{ 64 \ln \left( \frac{\varphi}{\Lambda } \right)^2 \left( \xi^{-1} + f \cdot \frac{3}{16} \right)}.
\end{align}
Inflation ends when $\epsilon =1$  such that:
\begin{align}
 \frac{\varphi_{\rm end}}{\Lambda} &= \exp \left( \frac{1}{8 \sqrt{\left(\xi ^{-1}  +f \cdot \frac{3}{16}\right)}} \right) \label{phiend}.
\end{align}
In the large field limit the number of e-foldings (\ref{efold}) is:
\begin{align}
N  
\simeq \left[  16 \left(\xi^{-1}+f \cdot \frac{3 }{16} \right) \ln \left( \frac{\varphi}{\Lambda}\right)^2 \right]_{\varphi_{end}}^{\varphi_{ini}}.
\end{align}
A simple way to determine the value of $\varphi_{ini}$ associated to when inflation starts is to require a minimal numbers of e-foldings compatible with a successful inflation, i.e. $N=60$. This leads to: 
\begin{align} 
\frac{\varphi_{ini}}{\Lambda} & \simeq \exp \left( \sqrt{\frac{60}{16 \left( \xi^{-1}+f \cdot \frac{3}{16} \right)} }\right) \ .
\end{align}
Further relevant information can be extracted using the  WMAP \cite{arXiv:0812.3622} normalization condition:
\begin{align}
\frac{U_{ini}}{\epsilon_{ini} } = (0.0276 M_{\rm P})^4.
\end{align}
The label {\it ini} signifies that this expression has to be evaluated at the beginning of the inflationary period. This condition helps estimating the magnitude of the non-minimal coupling. We deduce: 
\begin{align}
U_{ini}  
 \simeq \frac{2 M^{4}_{\rm P}}{\xi^2} \ln \left( \frac{\varphi_{ini}}{\Lambda} \right)  
\simeq \frac{2 M^{4}_{\rm P}}{\xi^2} \sqrt{\frac{3.75}{\xi^{-1}+ f \cdot 0.1875}}\ .
\end{align} 
while:
\begin{align}
\epsilon_{ini} \simeq \frac{1}{ 64 \ln \left( \frac{\varphi_{ini}}{\Lambda } \right)^2 \left( \xi^{-1} + f \cdot \frac{3}{16} \right)} = 0.0042\ .
\end{align}
We can therefore determine the magnitude of the non-minimal coupling which, depending whether we used the Palatini or the metric formulation, assumes the following value:  
\begin{align}
\xi  \simeq 1.4 \cdot 10^6 \quad \text{Palatini} \ , \qquad  {\rm and} \qquad 
\xi \simeq 6.1 \cdot 10^{4} \quad \text{Metric}
\end{align}
The knowledge of the non-minimal coupling allows us to estimate the initial and final value of the composite glueball field $\Phi$. We have in units of the strong scale $\Lambda$: 
\begin{align}
&\frac{\varphi_{end}}{\Lambda} \sim 10^{63.5}, \qquad \qquad  \frac{ \varphi_{ini}}{\Lambda} \sim 10^{986} \qquad \qquad \,\,\,{\rm Palatini.} \\&& \nonumber \\
&\frac{\varphi_{end}}{\Lambda} \sim 1.3 , \qquad \qquad \,\,\,\,\,\, \frac{\varphi_{ini}}{\Lambda}\sim 88 \qquad \qquad \qquad {\rm Metric .} 
\end{align}

From these results it is clear that the metric formulation directly provides a more natural range of values for $\varphi$ . Therefore, at the level of the present analysis and without invoking extra operators to match the metric and the Palatini formulation, we suggest to use the metric formulation when investigating/comparing strongly coupled inflationary models. {The effective action built here is a generating functional for trace anomaly and therefore the associated potential $V_{GI}$ cannot be quantum modified. This may protect the inflationary scenario even for large values of the scalar field.  Furthermore future first principle lattice simulations will be able to investigate the full nonperturbative physics.}

It is possible to further relate the strongly coupled scale $\Lambda$ with $\mathcal{M}$ recalling that we are working in the large field regime (\ref{lfe}). This implies that the  smallest value assumed by the inflaton must satisfy (\ref{lfe}) and therefore we obtain: 
\begin{equation}
\Lambda >\frac{\mathcal{M}}{\sqrt{\xi}} \qquad \qquad {\rm Metric} \ .
\end{equation}
$\mathcal{M}$ is the reduced Planck mass $2.44\cdot 10^{18}$ GeV yielding: 
\begin{equation}
 \Lambda > 0.9 \cdot 10^{16}\,\,\,{\rm GeV} \ .
\end{equation}
This is the typical scale for grand unification, in complete agreement with our earlier results for the first  model of composite inflation \cite{Channuie:2011rq}. One of the main differences with the model presented in \cite{Channuie:2011rq} is that here the full low energy potential of the inflaton is determined by matching trace anomaly between the underlying gauge theory and the effective action.  { As for the case of Higgs inflation, and other earlier approaches \cite{Spokoiny:1984bd,Futamase:1987ua,Salopek:1988qh,Fakir:1990eg,Kaiser:1994vs,Komatsu:1999mt,Tsujikawa:2004my} we discover that a phenomenologically large value of $\xi$ is needed for generating the correct size of the  observed amplitude of density fluctuations. A more complete treatment for all these models would require, in the future, a mechanism for generating such a large coupling.} 

\section{Glueball Inflation versus Unitarity}
\label{unitarity}
In this section we turn to the interesting question of the constraints set by tree-level unitarity of the inflaton field. For the present purpose it is convenient first to shift the overall Glueball potential, before coupling it non-minimally to gravity, in such a way that the potential evaluated on the ground state has zero energy:  
\begin{equation}
V_{\rm GI} \rightarrow 2\varphi^4\ln \left(\frac{\varphi}{\Lambda}\right) + \frac{\Lambda^4}{2 \,e} \ .
 \end{equation}
 The reason for such a shift is that in this case the ground state of the theory assumes the same value in the Jordan and in Einstein frame and reads: 
 \begin{equation}
 \langle \varphi \rangle = e^{-\frac{1}{4}}\Lambda = v\ .
 \end{equation}
The previous inflationary analysis remains unmodified by this shift. Furthermore we are interested in the large field expansion (\ref{lfe}) which can be well approximated by setting $\mathcal{M}=0$. The following relation is then natural: 
\begin{equation}
M_{\rm P}^2  \simeq  \xi v^2 \ , \qquad \Rightarrow \qquad \Omega = \frac{\varphi}{v} \ . 
\end{equation}

In the Einstein frame we then have: 
  \begin{equation}
\mathcal{S}_{\rm GI, \varphi}=\int d^{4}x \sqrt{-{g}}\left[-\frac{M_{\rm P}^{2}}{2}{g}^{\mu\nu}R_{\mu\nu}+ 16 \frac{v^2}{\varphi^2}\left(1+ \frac{3}{16}f\xi  \right)g^{\mu\nu}\, \partial_{\mu}\varphi \partial^{\nu}\varphi- \frac{v^4}{\varphi^4}\left[2\varphi^4 \ln\left(\frac{\varphi}{e^{\frac{1}{4}}v}\right) + \frac{v^4}{2}\right] \right]. \label{Gnonminimal}
\end{equation} 
We are now equipped with the needed ingredients to tackle the issue of tree-level unitarity at the effective Lagrangian level during the inflationary period. More specifically we are concerned with violation of tree-level unitarity of the scattering amplitude concerning the inflaton field fluctuations  $\delta \varphi$ around its classical time dependent background $\varphi_c(t)$ during the inflationary period. 
Following the analysis performed in \cite{Bezrukov:2010jz} we can, in first approximation, neglect the time dependence of the classical field and write: 
\begin{equation}
 \varphi = \varphi_c + \delta \varphi \ ,
 \end{equation}
 since the fluctuations are expected to encapsulate the high frequency modes of the inflaton. To estimate the actual cutoff of the tree-level scattering amplitude we analyze independently the kinetic and potential term for the inflaton in the Einstein frame. Starting from the kinetic term it is straightforward to show that around the classical background can be written as: 
 \begin{equation}
 \frac{v^2}{2\varphi_c^2}(32+6 f\xi) (\partial \delta \varphi)^2 
 \sum_{n=0}^{\infty} (n+1) \frac{(-\delta\varphi)^n}{\varphi_c^n} \ .
 \label{Kint}
 \end{equation}
It is possible to canonically normalize the first term of the series, i.e. the kinetic term for a free field, rescaling the fluctuations as follows:
\begin{equation}
\frac{ \delta \varphi}{ \varphi_c} = \frac {\delta \widetilde{\varphi} }{v\sqrt{32+6 f\xi }}\ . 
\end{equation} 
Under this field redefinition (\ref{Kint}) becomes:
\begin{equation}
\frac{(\partial \delta \widetilde{\varphi})^2}{2} \sum_{n=0}^{\infty} (n+1) \frac{(-\delta \widetilde{\varphi})^n}{ (32 + 6 f \xi)^{\frac{n}{2}}v^n} \ .
 \label{normKint}
\end{equation}
For the potential term  the higher order operators are also of the form: 
\begin{equation}
{\rm constant} \frac{(\delta \widetilde{\varphi})^n}{ (32 + 6 f \xi)^{\frac{n}{2}}v^n} \ .
\end{equation}
This implies that the tree-level cutoff for unitarity is in the metric formulation: 
\begin{equation}
\sqrt{\xi} v \simeq M_{\rm P}  
\end{equation}
while it is simply $v$ in the Palatini formulation. 
This results shows that the cutoff, in both formulations, is background independent. Quite nicely the unitarity cutoff in the metric formulation corresponds to the Planck scale and therefore tree-level unitarity is safe in this approach, however this is not the case for the Palatini formulation. These results are in complete agreement with the findings for successful inflation in the previous section. 

\section{Graviton exchange for Composite Inflation} 
\label{graviton-scattering}

Similar to the case of Higgs inflation, composite inflation introduces a non-minimal coupling to gravity of the type $\xi \varphi^2 R$ allowed by all known symmetries of the underlying strongly coupled theory and gravity. In \cite{Burgess:2009ea,Barbon:2009ya,Burgess:2010zq,DeSimone:2008ei} it is argued that, although this term superficially appears to be a dimension four operator, expanding it around flat space, $g_{\mu \nu} = \eta_{\mu \nu} + h_{\mu \nu}/ M_{\rm P}$, leads to a dimension five operator plus an infinite tower of higher dimensional operators: 
\begin{equation}
\xi \varphi^2 R \sim \xi \varphi^2   \frac{\square h}{M_{\rm P}} + \dots \ .
\end{equation}
This indicates that  generic non-minimally coupled theories become strongly interacting at scales $\Lambda_{\rm NRG} \sim M_{\rm P}/\xi$. The new scale $\Lambda_{\rm NRG}$ emerges because gravity in four dimensions is non renormalizable and NRG stands for Non Renormalizable Gravity. In the case of minimally coupled theories, this scale is simply $M_{\rm P}$. Therefore, without any protecting mechanism, the interaction with gravity can lead to a series of corrections to the 
low energy effective Lagrangian. Using the canonically normalized field $\varphi$, one naively expects the following corrections to any potential, and in our specific case to $V_{\rm GI}$: 
\begin{equation}
V = V_{\rm GI}(\Lambda) +  \varphi^4 \sum_{n>0} a_{n} \left( \frac{\varphi}{\Lambda_{\rm NRG}}\right)^n + \xi \varphi^2 R \sum_{n} b_{n}  \left( \frac{\varphi}{\Lambda_{\rm NRG}}\right)^n \ .
 \end{equation}
The new interactions are suppressed by $\Lambda_{\rm NRG} \sim M_{\rm P}/\xi$ while the new strongly coupled dynamics has a scale $\Lambda \sim M_{\rm P}/\sqrt{\xi}$.  The coefficients $a_n$ and $b_n$, due to graviton exchange, depend on the behavior of gravity above the scale $\Lambda_{\rm NRG}$. Unless a protecting mechanism exists, and taking all the coefficients $a_n$ and $b_n$ to be of order unity,  the flatness, in the Einstein frame, of the inflationary potential can be questioned. This is not only the case of Higgs inflation, but also of many minimal models of inflation, such as $m^2\varphi^2$ chaotic inflation, since in these cases $\varphi > \Lambda_{\rm NRG}$ during inflation. 

Although no actual resolution to this potential issue was presented in \cite{Burgess:2009ea,Barbon:2009ya,Burgess:2010zq,DeSimone:2008ei}, it was, however, pointed out that currently we have no experimental evidence that $a_n$ and $b_n$ must be of order unity and that there is still the logical possibility that graviton exchange is softer than naive estimate suggested in \cite{Horava:2009uw} leaving our potential unaltered.  We could therefore work in the same spirit of Higgs or chaotic inflation with the further benefit that, as we showed above, the inflaton-inflaton scattering is better behaved than in models of Higgs inflation. 

In composite inflation, there is already a symmetry principle partially constraining the effective potential $V_{\rm GI}$. This constraint requires the action for $\varphi$ to be such that, at zero external momentum, the matter trace-anomaly, in the Jordan frame, has to reproduce the Yang-Mills trace anomaly and therefore automatically requires $a_n = 0$ for any $n>0$. The situation for the $b_n$ coefficients is more delicate since they involve derivative vanishing at zero momentum, however, it would seem natural that also these coefficients have to vanish.

\section{Summary of the different energy scales}
\label{scales}
For the benefit of the reader we summarize the various scales and associated operators involved in the present setup before and after coupling our underlying gauge theory to gravity. 

We started our exploration by introducing the {\it simplest} non-abelian gauge theory known, i.e. the pure $SU(N)$ Yang-Mills gauge theory. The fundamental Lagrangian for this gauge theory, in absence of the $\theta$-angle operator, is constituted by {\it only one} renormalizable conformal operator\footnote{If we add also the $\theta$-angle operator we have one more renormalizable conformal operator which does not affect the classical equations of motion.}: 
\begin{equation} 
\mathcal{L}_{\rm Fund} = -\frac{1}{4} \sum_{a=1}^{N^2} \mathcal{G}^{\mu \nu}_{a} \mathcal{G}_{\mu \nu,a} \ . \label{fundamental}
\end{equation}
First principle lattice simulations have shown that this theory confines and via dimensional transmutation a renormalization invariant physical scale is generated. This scale is identifiable with the scale $\Lambda$ of the glueball theory introduced in the previous sections. Using the renormalization group equations, lattice simulations, as well as our experience from ordinary quantum chromodynamics\footnote{Which is Yang-Mills with quarks.}  the fundamental theory can be used in the perturbative regime to describe the dynamics of the theory at energy scales of the order of $100\,\Lambda$ and above. For energies below this scale and to describe the vacuum properties of the theory the effective potential given in (\ref{effective}) works and it has been used recently in \cite{Campbell:2011iw} also to determine cosmological properties. 

When coupling our theory to gravity we can, of course, use directly the unique operator constituting the fundamental gauge theory (\ref{fundamental}), and use, for example first principle lattice simulations. However, because we were interested in slow roll conditions near the ground state of the underlying gauge theory we used the simplest and most appropriate analytic description, i.e. the one in terms of the glueball effective theory. As an important consistency check we showed that inflation starts at energy scales just below or near the energy scales above which the underlying gauge dynamics is perturbative and described by a single renormalizable operator. We have also showed that the natural scale for $\Lambda$ is the grand-unified scale which is orders of magnitude smaller than the Planck scale. Therefore we expect the perturbative dynamics of the gauge theory to set-in before we arrive at the Planck scale. We showed, furthermore, that inflaton-inflaton scattering would only be affected by Planck scale physics making our analysis, from this point of view, more solid than Higgs inflation. 

{ The grand-unified scale here is defined as the energy at which the standard model gauge couplings, in a given renormalisation scheme, unify. Given that the standard model alone does not unify, an extension perhaps also including dark matter  is needed. The standard model couplings are weak at the unification point. However the inflationary model is still strongly coupled at this scale (now identified with $\Lambda$). Therefore, a potential unification of the standard model and the new inflationary gauge dynamics can only take place at or around the Planck scale  which is not accessible with our current understanding of the gravitational corrections.} 

There is, however, another scale to worry about, i.e. the one associated to graviton scattering. In the last section we have shown that, like in Higgs inflation and several other scenarios, this problem arises at a new scale $\Lambda_{\rm NRG}<\Lambda$. The fact that this scale $\Lambda_{\rm NRG}$ is smaller than $\Lambda$, i.e. where inflation takes place, might spoil the inflationary scenario unless a mechanism for softening this behavior emerges. Due to the fact that this mechanism, as stressed above \cite{DeSimone:2008ei}, must be active above the scale $\Lambda_{\rm NRG}$ this implies the following scenarios for composite inflation. If the scale where this mechanism emerges is below $10$  to $100\,\Lambda$ then the effective description given in (\ref{effective}) is valid and  we can use the further constraint $a_n=0$ needed to correctly saturate the trace anomaly of the underlying gauge theory. If the mechanism is introduced at scales between $100\Lambda$ and $M_{\rm P}$ the underlying Lagrangian, before coupling to gravity, reduces to (\ref{fundamental}). In this energy range the underlying gauge theory is perturbative and therefore one can use any mechanism that works for Higgs inflation. Finally, if the scale at which this mechanism takes place is above $M_{\rm P}$ a more complete theory of gravity is needed. This shows that our model has, in the worse case scenario, the same limitations of Higgs inflation for graviton scattering but works better for inflaton scattering.
 
 \section{Conclusions}
\label{conclusions}

We further investigated the paradigm according to which inflation is driven by a four-dimensional strongly coupled dynamics non-minimally coupled to gravity. We have done so by introducing an explicit model where the inflaton is identified with the glueball field of a pure Yang-Mills theory.  We used the well known dilatonic-like glueball action. This model constitutes the building block of any model of composite inflation. We showed that successful inflation can be achieved. Furthermore the confining scale of the Yang-Mills theory, for a successful inflation, matches the one of the grand unified energy scale. This result is in line with the result found in \cite{Channuie:2011rq}. We discovered that within the metric formulation models of composite inflation lead to a more consistent picture than within the Palatini one.  Another welcome feature of glueball inflation, in the metric formulation, is that we found the model to respect tree-level unitarity, for the scattering of the inflaton field during inflation, all the way to the Planck scale. Furthermore using the knowledge of the phase diagram of strongly coupled theories \cite{Sannino:2004qp,Dietrich:2006cm,Ryttov:2007sr,Ryttov:2007cx,Pica:2010mt,Sannino:2009aw,Mojaza:2010cm,Pica:2010xq} we can, in the future, explore several dynamical models of inflation. 

 \acknowledgments
The work of F.B. is partially supported by the Humboldt foundation, while the one of P.C. is supported by the Royal Thai Government under the program {\it Strategic Scholarships for Frontier Research Network} of the Thailand's  Commission for Higher Education.

\appendix 
\section{Composite Inflation Setup}
\label{CISET}
We consider a generic strongly coupled theory before coupling it to gravity. We identify the inflaton with one of the lightest composite states of the theory and denote it with $\Phi$. This state has mass dimension $d$. This is the physical dimension coming from the sum of the engineering dimensions of the elementary fields constituting the inflaton augmented by the anomalous dimensions due to quantum corrections in the underlying gauge theory. {}In this work we concentrate on the non-Goldstone sector of the theory\footnote{The Goldstone sector, if any,  associated to the potential dynamical spontaneous breaking of some global symmetries of the underlying gauge theory will be investigated elsewhere}. 

We consider the following coupling to gravity in the Jordan frame: 
\begin{align}
\mathcal{S}_{\rm CI,J}=\int d^{4}x \sqrt{-g}\left[-\frac{\mathcal{M}^{2}+\xi\,{\Phi}^{\frac{2}{d}}}{2}g^{\mu\nu}R_{\mu\nu}+\mathcal{L}_{\Phi}\right], \quad\mathcal{L}_{\Phi}=g^{\mu\nu}\Phi^{\frac{2-2d}{d}}\partial_{\mu}\Phi\partial_{\nu}\Phi-V({\Phi}), \label{nonminimal}
\end{align}
with $\mathcal{L}_{\Phi}$ the low energy effective Lagrangian for the field $\Phi$ constrained by the symmetries of the underlying strongly coupled theory.  In this framework $\mathcal{M}$ is not automatically the Planck constant $M_{\rm Pl}$. The non-minimal coupling to gravity is controlled by the dimensionless coupling $\xi$. The non-analytic power of $\Phi$  emerges because we are requiring a dimensionless coupling with the Ricci scalar. Abandoning the conformality requirement allows for operators with integer powers of $\Phi$ when coupling to the Ricci scalar. However a new energy scale must be introduced to match the mass dimensions. 

We diagonalize the gravity-composite dynamics model via the conformal transformation: \begin{align}
g_{\mu\nu}\rightarrow\tilde{g}_{\mu\nu}=\Omega({\Phi})^2 g_{\mu\nu},\quad\Omega({\Phi})^2=\frac{\mathcal{M}^2+\xi\Phi^{\frac{2}{d}}}{M_{\rm P}^2},
\end{align}
such that 
\begin{align}
\quad\tilde{g}^{\mu \nu}=\Omega^{-2}g^{\mu\nu},\quad\sqrt{-\tilde{g}}=\Omega^4\sqrt{-g}.
\end{align}
 We use both the Palatini and the metric formulation. The difference between the two formulations resides in the fact that in the Palatini formulation the connection $\Gamma$ is assumed not to be directly associated with the metric $g_{\mu\nu}$. Hence the Ricci tensor $R_{\mu\nu}$ does not transform under the conformal transformation.  
 
Applying the conformal transformation we land in the Einstein frame and the action reads:
\begin{align}
\mathcal{S}_{\rm CI,E} =\int d^{4}x \sqrt{-g}\left[ -\frac{1}{2} M_{\rm P}^2 \,\, g^{\mu \nu}R_{\mu \nu} + \Omega^{-2} \left(\Phi^{\frac{2-2d}{d}} + f \cdot 3 M^{2}_{\rm P} {\Omega'} ^{2} \right)g^{\mu \nu} \partial_{\mu} \Phi \partial_{\nu} \Phi  - \Omega ^{-4} V({\Phi}) \right].
\end{align}
Primes denotes derivatives with respect to $\Phi$ and tildes are dropped for convenience. $f=1$ signifies the metric formulation \cite{Kaiser:1994vs,Tsujikawa:2000wc,Bezrukov:2008ut,Barvinsky:2008ia} and $f=0$ the Palatini one \cite{Bauer:2010jg}.  

We landed with an involved kinetic term for the inflaton.  It is convenient to introduce a canonically normalized field $\chi$ related to $\Phi$ via
\begin{align}
\frac{1}{2} \tilde{g}^{\mu \nu} \partial_{\mu} \chi (\Phi) \partial_{\nu} \chi(\Phi) = \frac{1}{2} \left( \frac{d \chi}{d \Phi} \right)^2 \tilde{g}^{\mu \nu} \partial_{\mu} \Phi \partial_{\nu} \Phi \ ,
\end{align}
with
\begin{align}
\frac{1}{2} \left( \frac{d \chi}{d \Phi} \right)^2 &= \Omega^{-2} \left(\Phi ^{\frac{2-2d}{d}} + f \cdot 3 M^{2}_{\rm P} {\Omega'} ^2 \right) = \Omega^{-2}\left(1 +  f \cdot\frac{3 \xi ^2}{d^2 M_{\rm P}^2} \Omega^{-2} \Phi^{\frac{2}{d}} \right) \Phi ^{\frac{2-2d}{d}}. \label{defchi}
\end{align}
In terms of the canonically normalized field we have: 
\begin{align}
\mathcal{S}_{\rm CI,E} &=\int d^{4}x \sqrt{-g}\left[-\frac{1}{2} M_{\rm P}^2 g^{\mu \nu}R_{\mu \nu} + \frac{1}{2} g^{\mu \nu} \partial_{\mu} \chi \partial_{\nu} \chi- U(\chi)  \right].
\end{align}

With
\begin{align}
U(\chi) \equiv \Omega^{-4}V(\Phi).  
\end{align}

We will analyze the dynamics in the Einstein frame, and therefore define the slow-roll parameters in terms of $U$ and $\chi$:
\begin{align}
\epsilon = \frac{M_{\rm P}^2}{2} \left( \frac{dU / d \chi}{U} \right)^2, \quad \quad \eta = M_{\rm P}^2 \left( \frac{d^2U / d \chi^2}{U} \right), \quad \quad N = \frac{1}{M_{\rm P}^2} \int _{\chi_{end}} ^{\chi_{ini}} \frac{U}{dU /d\chi} d \chi. \label{epsilon}
\end{align}
We will, however, express everything in terms of $\Phi$, such that we don't need an explicit solution of (\ref{defchi}). We obtain:
\begin{align}
\epsilon = \frac{M_{\rm P}^2}{2}  \left( -4 \Omega^{-1} \Omega' + \frac{V'}{V}\right) ^2 \left( \frac{1}{{\chi}'}\right)^2 = \frac{1}{4} \frac{\left( \left( 1+ \frac{\xi}{{\cal{M}}^2} \Phi^{\frac{2}{d}}\right)\Phi \frac{V'}{V}- \frac{4}{d} \frac{\xi}{\mathcal{M}^2} \Phi^{\frac{2}{d}} \right)^2}{\left(1 + \frac{\xi}{\mathcal{M}^2} \Phi^{\frac{2}{d}} \right)\ \frac{1}{\mathcal{M}^2}\Phi^{\frac{2}{d}} + f \cdot \frac{3}{d^2}\left( \frac{\xi}{\mathcal{M}^2} \Phi^{\frac{2}{d}}\right)^2}\ ,
\end{align}

\begin{align}
\eta = M_{\rm P}^2 \left( \frac{V'' {\chi}' - V' {\chi}''+20 \Omega^{-2} \left( \Omega '\right)^2 V -4 \Omega^{-1}  \Omega '' V  -8 \Omega^{-1} \Omega ' V' +4\Omega^{-3} \Omega' {\chi}'' V}{V {\chi'}^3}\right)\ ,
\end{align}

\begin{align}
N = \frac{1}{M_{\rm P}^2} \int _{\Phi_{end}} ^{\Phi_{ini}} \frac{V}{-4 \Omega^{-1}\Omega ' V + V'} {\chi'}^2 d \Phi = \frac{2}{\mathcal{M}^2}\int_{\Phi_{end}}^{\Phi_{ini}} \frac{\Phi^{\frac{2-d}{d}}\left(1+f \cdot \frac{3 \xi^2}{d^2 \mathcal{M}^2} \Phi^{\frac{2}{d}} \frac{1}{1+ \frac{\xi}{\mathcal{M}^2}\Phi^{\frac{2}{d}}}\right)}{\frac{-4}{d}\frac{\xi}{\mathcal{M}^2}\Phi^{\frac{2}{d}}+ \left(1+ \frac{\xi}{\mathcal{M}^2}\Phi^{\frac{2}{d}} \right)\Phi\frac{V'}{V}} d \Phi \ . \label{efold}
\end{align}
 
Our framework resembles the one of Higgs-inflation  \cite{Bezrukov:2007ep} with the difference that our inflaton stems from a natural four-dimensional dynamics and therefore it is free from unnatural fine-tuning. Of course, as for the Higgs-inflation paradigm \footnote{Here it was proposed that the inflationary expansion of the early Universe can be linked to the standard model by identifying the standard model Higgs boson with the inflaton. The salient feature of the Higgs-inflation mechanism is the non-minimal coupling of the Higgs  doublet field ($H$) to gravity. This happens by adding a term of the type $\xi H^{\dagger} H R$ to the standard gravity-matter action, with $\xi$ a new coupling constant. This non-minimal coupling of scalar fields to gravity has a long history \cite{Spokoiny:1984bd,Futamase:1987ua,Salopek:1988qh,Fakir:1990eg,Kaiser:1994vs,Komatsu:1999mt,Tsujikawa:2004my}. A nonzero value of $\xi$ is needed since for $\xi=0$ an unacceptably large amplitude of primordial inhomogeneities is generated for a realistic quartic Higgs self-interaction term \cite{Linde:1983gd}. It was found in \cite{Bezrukov:2007ep} that with $\xi$ of the order $10^4$ the model leads to successful inflation, provides the graceful exit from it, and produces the spectrum of primordial fluctuations in good agreement with the observational data. This scenario was further explored in \cite{Barvinsky:2008ia,Bezrukov:2008ut,GarciaBellido:2008ab,DeSimone:2008ei,Bezrukov:2008ej,Bezrukov:2009db,Barvinsky:2009fy,Burgess:2009ea,Barbon:2009ya,Burgess:2010zq,Atkins:2010yg,Bezrukov:2010jz}. }, the composite inflation framework still begs for an explanation of the non-minimal coupling to gravity. We set aside this important point in this initial investigations but point out that a first glimpse of how the associated operator might be viewed from a more elementary point of view has been briefly discussed in \cite{Channuie:2011rq}. 
  
Our framework, based on generic four-dimensional strongly coupled gauge theories, constitutes the natural template for other models of composite inflation using, for example, holographic inspired descriptions of strongly coupled dynamics \cite{Evans:2010tf,Chen:2010pd}.

\end{document}